%
\newcount\mgnf\newcount\tipi\newcount\tipoformule
\newcount\aux\newcount\piepagina\newcount\xdata
\mgnf=0
\aux=0           
\tipoformule=1   
\piepagina=1     
\xdata=1         
\def\Di{16\, Mar\, 1997}

\ifnum\mgnf=1 \aux=0 \tipoformule =1 \piepagina=1 \xdata=1\fi
\newcount\bibl
\ifnum\mgnf=0\bibl=0\else\bibl=1\fi

%
%
%
%
\openout8=ref.b
\ifnum\bibl=0
\def\ref#1#2#3{[#1#2]\write8{#1@#2}}
\def\rif#1#2#3#4{\item{[#1#2]} #3}
\fi

\ifnum\bibl=1
\def\ref#1#2#3{[#3]\write8{#1@#2}}
\def\rif#1#2#3#4{}

\fi

\def\9#1{\ifnum\aux=1#1\else\relax\fi}
\ifnum\piepagina=0 \footline={\rlap{\hbox{\copy200}\
$\st[\number\pageno]$}\hss\tenrm \foglio\hss}\fi \ifnum\piepagina=1
\footline={\rlap{\hbox{\copy200}} \hss\tenrm \folio\hss}\fi
\ifnum\piepagina=2\footline{\hss\tenrm\folio\hss}\fi

\ifnum\mgnf=0 \magnification=\magstep0
\hsize=12truecm\vsize=19.5truecm \parindent=4.pt\fi
\ifnum\mgnf=1 \magnification=\magstep1
\hsize=16.0truecm\vsize=22.5truecm\baselineskip14pt\vglue5.0truecm
\overfullrule=0pt \parindent=4.pt\fi

\let\a=\alpha   
\let\e=\varepsilon \let\z=\zeta 
 \let\l=\lambda \let\m=\mu 
\let\x=\xi \let\p=\pi  \let\s=\sigma \let\t=\tau
 \let\f=\varphi  
  \let\D=\Delta 
\let\L=\Lambda   
  
{\count255=\time\divide\count255 by 60 \xdef\oramin{\number\count255}
\multiply\count255 by-60\advance\count255 by\time
\xdef\oramin{\oramin:\ifnum\count255<10 0\fi\the\count255}}
\def\ora{\oramin }

\ifnum\xdata=0
\def\data{\number\day/\ifcase\month\or gennaio \or
febbraio \or marzo \or aprile \or maggio \or giugno \or luglio \or
agosto \or settembre \or ottobre \or novembre \or dicembre
\fi/\number\year;\ \ora}
\else
\def\data{\Di}
\fi

\setbox200\hbox{$\scriptscriptstyle \data $}
\newcount\pgn \pgn=1
\def\foglio{\number\numsec:\number\pgn
\global\advance\pgn by 1} \def\foglioa{A\number\numsec:\number\pgn
\global\advance\pgn by 1}
\global\newcount\numsec\global\newcount\numfor \global\newcount\numfig
\gdef\profonditastruttura{\dp\strutbox}
\def\senondefinito#1{\expandafter\ifx\csname#1\endcsname\relax}
\def\SIA #1,#2,#3 {\senondefinito{#1#2} \expandafter\xdef\csname
#1#2\endcsname{#3} \else \write16{???? ma #1,#2 e' gia' stato definito
!!!!} \fi} \def\etichetta(#1){(\veroparagrafo.\veraformula) \SIA
e,#1,(\veroparagrafo.\veraformula) \global\advance\numfor by 1
\9{\write15{\string\FU (#1){\equ(#1)}}} \9{ \write16{ EQ \equ(#1) == #1
}}} \def \FU(#1)#2{\SIA fu,#1,#2 }
\def\etichettaa(#1){(A\veroparagrafo.\veraformula) \SIA
e,#1,(A\veroparagrafo.\veraformula) \global\advance\numfor by 1
\9{\write15{\string\FU (#1){\equ(#1)}}} \9{ \write16{ EQ \equ(#1) == #1
}}} \def\getichetta(#1){Fig.  \verafigura \SIA e,#1,{\verafigura}
\global\advance\numfig by 1 \9{\write15{\string\FU (#1){\equ(#1)}}} \9{
\write16{ Fig.  \equ(#1) ha simbolo #1 }}} \newdimen\gwidth \def\BOZZA{
\def\alato(##1){ {\vtop to \profonditastruttura{\baselineskip
\profonditastruttura\vss
\rlap{\kern-\hsize\kern-1.2truecm{$\scriptstyle##1$}}}}}
\def\galato(##1){ \gwidth=\hsize \divide\gwidth by 2 {\vtop to
\profonditastruttura{\baselineskip \profonditastruttura\vss
\rlap{\kern-\gwidth\kern-1.2truecm{$\scriptstyle##1$}}}}} }
\def\alato(#1){} \def\galato(#1){}
\def\veroparagrafo{\number\numsec}\def\veraformula{\number\numfor}
\def\verafigura{\number\numfig}
\def\geq(#1){\getichetta(#1)\galato(#1)}
\def\Eq(#1){\eqno{\etichetta(#1)\alato(#1)}}
\def\eq(#1){\etichetta(#1)\alato(#1)}
\def\Eqa(#1){\eqno{\etichettaa(#1)\alato(#1)}}
\def\eqa(#1){\etichettaa(#1)\alato(#1)}
\def\eqv(#1){\senondefinito{fu#1}$\clubsuit$#1\write16{No translation
for #1} \else\csname fu#1\endcsname\fi}
\def\equ(#1){\senondefinito{e#1}\eqv(#1)\else\csname e#1\endcsname\fi}
\openin13=#1.aux \ifeof13 \relax \else \input #1.aux \closein13\fi
\openin14=\jobname.aux \ifeof14 \relax \else \input \jobname.aux
\closein14 \fi \9{\openout15=\jobname.aux} \newskip\ttglue


\font\titolo=cmbx10 scaled \magstep1
\font\ottorm=cmr7\font\ottoi=cmmi7\font\ottosy=cmsy7
\font\ottobf=cmbx7\font\ottott=cmtt8\font\ottosl=cmsl8\font\ottoit=cmti7
\font\sixrm=cmr7\font\sixbf=cmbx7\font\sixi=cmmi7\font\sixsy=cmsy7

\font\fiverm=cmr5\font\fivesy=cmsy5\font\fivei=cmmi5\font\fivebf=cmbx5
\def\ottopunti{\def\rm{\fam0\ottorm}\textfont0=\ottorm%
\scriptfont0=\sixrm\scriptscriptfont0=\fiverm\textfont1=\ottoi%
\scriptfont1=\sixi\scriptscriptfont1=\fivei\textfont2=\ottosy%
\scriptfont2=\sixsy\scriptscriptfont2=\fivesy\textfont3=\tenex%
\scriptfont3=\tenex\scriptscriptfont3=\tenex\textfont\itfam=\ottoit%
\def\it{\fam\itfam\ottoit}\textfont\slfam=\ottosl%
\def\sl{\fam\slfam\ottosl}\textfont\ttfam=\ottott%
\def\tt{\fam\ttfam\ottott}\textfont\bffam=\ottobf%
\scriptfont\bffam=\sixbf\scriptscriptfont\bffam=\fivebf%
\def\bf{\fam\bffam\ottobf}\tt\ttglue=.5em plus.25em minus.15em%
\setbox\strutbox=\hbox{\vrule height7pt depth2pt width0pt}%
\normalbaselineskip=9pt\let\sc=\sixrm\normalbaselines\rm}
\catcode`@=11
\def\footnote#1{\edef\@sf{\spacefactor\the\spacefactor}#1\@sf
\insert\footins\bgroup\ottopunti\interlinepenalty100\let\par=\endgraf
\leftskip=0pt \rightskip=0pt \splittopskip=10pt plus 1pt minus 1pt
\floatingpenalty=20000
\smallskip\item{#1}\bgroup\strut\aftergroup\@foot\let\next}
\skip\footins=12pt plus 2pt minus 4pt\dimen\footins=30pc\catcode`@=12
\newdimen\xshift \newdimen\xwidth \newdimen\yshift
\def\ins#1#2#3{\vbox to0pt{\kern-#2 \hbox{\kern#1
#3}\vss}\nointerlineskip} \def\eqfig#1#2#3#4#5{ \par\xwidth=#1
\xshift=\hsize \advance\xshift by-\xwidth \divide\xshift by 2
\yshift=#2 \divide\yshift by 2 \line{\hglue\xshift \vbox to #2{\vfil #3
\includegraphics{#4.ps} }\hfill\raise\yshift\hbox{#5}}} \def\8{\write13}

\def\V#1{{\,\underline#1\,}}
\def\T#1{#1\kern-4pt\lower9pt\hbox{$\widetilde{}$}\kern4pt{}}
\let\dpr=\partial \let\io=\infty\let\ig=\int
\def\fra#1#2{{#1\over#2}}\def\media#1{\langle{#1}\rangle}\let\0=\noindent
\def\guida{\leaders\hbox to 1em{\hss.\hss}\hfill}
\def\tende#1{\vtop{\ialign{##\crcr\rightarrowfill\crcr
\noalign{\kern-1pt\nointerlineskip} \hglue3.pt${\scriptstyle
#1}$\hglue3.pt\crcr}}} \def\otto{\
{\kern-1.truept\leftarrow\kern-5.truept\to\kern-1.truept}\ }

\def\pagina{\vfill\eject}

\def\st{\scriptscriptstyle}
\def\*{\vskip0.3truecm}

\def\lis#1{{\overline #1}}\def\eg{\hbox{\it e.g.\ }}

\def\ie{\hbox{\it i.e.\ }}

\def\fiat{{}}
\def\\{\hfill\break} \def\={{ \; \equiv \; }}

\def\annota#1{\footnote{${}^#1$}}
\ifnum\aux=1\BOZZA\else\relax\fi
\ifnum\tipoformule=1\let\Eq=\eqno\def\eq{}\let\Eqa=\eqno\def\eqa{}
\def\equ{{}}\fi
\def\defi{\,{\buildrel def \over =}\,}
\def\1{\ifnum\mgnf=0\pagina\else\relax\fi}
\def\W#1{#1_{\kern-3pt\lower6.6truept\hbox to 1.1truemm
{$\widetilde{}$\hfill}}\kern2pt\,}

\def\FINE{
\*
\0{\it Internet access:
All Authors' quoted preprints can be also found and freely downloaded
(latest postscript version including misprints/errors corrections) at
the (mirror) sites:

\centerline{\tt http://chimera.roma1.infn.it}
\centerline{\tt http://www.math.rutgers.edu/$\sim$giovanni}

\0in the Mathematical Physics Preprints page.\\
\sl e-mail: giovanni@ipparco.roma1.infn.it
}}

\def\GG{{\V G}}

\def\NN{{\cal N}}\def\CC{{\cal C}}\def\EE{{\cal E}}
\def\FF{{\cal F}}

\ifnum\bibl=1
\def\diecipunti{................................}
\message{Compilare due volte per avere i rifermenti numerici}
\message{............\diecipunti.................................}

\fi

\fiat
\centerline{\titolo Fluctuation patterns and conditional reversibility}
\centerline{\titolo in nonequilibrium systems}
\*\*

\centerline{\bf Giovanni Gallavotti}
\*
\centerline{\it I.H.E.S., 91440 Bures s/Yvette, France}

\vskip1.truecm
{\bf Abstract:} {\it Fluctuations of observables as functions of time,
or {\sl fluctuation patterns}, are studied in a chaotic
microscopically reversible system that has irreversibly reached a
nonequilibrium stationary state.  Supposing that during a certain,
long enough, time interval the average entropy creation rate has a
value $s$ and that during another time interval of the same length it
has value $-s$ then we show that the relative probabilities of
fluctuation patterns in the first time interval are the same as those
of the reversed patterns in the second time interval.  The system is
``conditionally reversible'' or irreversibility in a reversible system
is "driven" by the entropy creation: while a very rare fluctuation
happens to change the sign of the entropy creation rate it also
happens that the time reversed fluctuations of all other observables
acquire the same relative probability of the corresponding
fluctuations in presence of normal entropy creation. A mathematical
proof is sketched.}

\vskip1truecm
\0{\it\S1. Entropy generation.}
\numsec=1\numfor=1 \*

We consider a reversible mechanical system governed by a smooth
equation:

$$\dot x=f(x ,\GG)\Eq(1.1)$$
depending on several parameters $\GG=(G_1,\ldots, G_n)$ measuring the
strength of the forces acting on the system and causing the evolution
$x \to S_tx $ of the phase space point $x$
representing the system state in the {\it phase space} $\FF$ which can
be, quite generally, a smooth manifold.

We suppose that the system is ``thermostatted'' so that motions take
place on bounded smooth invariant surfaces $H(x ;\GG)=E$, which are
level surfaces of some ``level function'' $H$.  Hence we shall
identify, to simplify the notations, the manifold $\FF$ with this
level surface.

We suppose that the flow $S_t$ generated by \equ(1.1) is {\it
reversible}, \ie there is a volume preserving smooth map $I$, ``time
reversal'', of phase space "anticommuting with time" and such that
$I^2=1$:

$$S_t I=iI S_{-t}\Eq(1.2)$$
\ie $f(I x )=-(\dpr I)(x )\cdot f(x )$.  If $-\s(x ;\GG)$ is the
rate of change of the volume element of $\FF$ near $x $ and under the
flow $S_t$ (it would be equal to $\sum_\a\dpr_\a f_\a(x ;\GG)$
if the space $\FF$ was a euclidean space) we shall call $\s$ the {\it
entropy generation rate} and we suppose that it has the properties:

$$\s(Ix ;\GG)=-\s(x ;\GG), \qquad \s(x ;\V0)=0\Eq(1.3)$$
so that at zero forcing the evolution is volume preserving (a property
usually true because the non forced system is very often Hamiltonian).

Our analysis concerns idealized systems of the above type
that are also transitive {\it Anosov flows} on each energy surface.

We recall that a Anosov flow $S_t$ is a flow, without equilibrium
points, solving a smooth differential equation $\dot{{x}}=f(x)$1 on a
smooth compact manifold; the flow is such that at every point $y$ one
can define a {\it stable tangent plane} $T^s_y$, and an {\it unstable
tangent plane} $T^u_y$ with:\\
1) the tangent planes $T^u_y$ and $T^s_y$ are linearly independent and
vary continuously with $y$ and $T^s_y,T^u_y$ togheter with the vector
$f(y)$ span the full tangent plane at $y$.\\
2) they are {\it covariant}: $\dpr S_t T^\a_y=T^\a_{S_ty}$, $\a=u,s$\\
3) there exist constants $C,\l>0$ such that
if $\x\in T^s_y$ then $|\dpr S_t \x|\le C e^{-\l t}|\x|$ for all
$t\ge0$, and
if $\x\in T^u_y$ then $|\dpr S_{-t} \x|\le C e^{-\l t}|\x|$ for all
$t\ge0$.

A Anosov flow is {\it mixing} or {\it transitive} if given any two open
sets $A,B$ there is a $t_{A,B}$ such that $S_t A\cap B\ne \emptyset$
for $|t|> t_{A,B}$.

It follows, see \ref{BR}{}{1}, that there exists a unique probability
distribution $\m$ on $\FF$ such that almost all initial data with
respect to the volume generate motions that ``admit a statistics
$\m$'', \ie:

$$\lim_{T\to\io}\fra1T\ig_0^T F(S_tx )dt=\ig \m(d y ) F( y ){\buildrel
def\over =}
\media{F}_+\Eq(1.4)$$
and $\m$ is called the {\it SRB distribution}.  The average of $F$ with
respect to $\m$ will be denoted $\media{F}_+$.

We shall further restrict our attention to transitive Anosov systems
that are reversible, in the above sense, for all values of the forcing
parameters $\V G$ of interest and {\it dissipative} at $\GG\ne\V0$.
This means that the systems we consider are such that:

$$\media{\s}_+>0 \qquad {\rm for}\ \GG\ne\V0\Eq(1.5)$$
Under the above assumptions one can define, for $\media{\s}_+>0$, the
``dimensionless average entropy creation rate'' $p$ by setting:

$$p=\fra1{\media{\s}_+\t}\ig_{-\t/2}^{\t/2} \s(S_t;\GG) dt\Eq(1.6)$$

Then the probability distribution of the variable $p$ with respect to
the SRB distribution $\m$ can be written for large $\t$ as
$\p_\t(p)dp=e^{-\t\z_\t(p)} dp$, see \ref{G}{6}{2}, and the function
$\z(p)=\lim_{\t\to\io}\z_\t(p)$ verifies, if $\s_+\=\media{\s}_+>0$ and
$|p|\le p^*$ for a suitable $p^*\ge1$, the property:

$$\fra{\z(-p)-\z(p)}{p\media{\s}_+}=1\Eq(1.7)$$
which is called a {\it fluctuation theorem}, and is part of a class of
theorems proved in \ref{GC}{}{3} for discrete time systems, and in
\ref{Ge}{}{4}, for continuous time systems.  This theorem can be considerably
extended, as discussed in \ref{G}{2}{5} and the extension can be shown
to imply, in the limit $\V G\to\V0$ (when also $\s_+\to0$) relations
that can be identified in various cases with Green--Kubo's formulae
and Onsager's reciprocal relations, see also \ref{GR}{}{6}.
\*

{\it The connection with applications of the above results is made via the
assumption that concrete chaotic dynamical systems can be considered,
"for the purpose of studying the properties of interest", as transitive
Anosov flows.  This assumption was called in \ref{GC}{}{3} the {\sl
chaotic hypothesis} and makes the above results of immediate relevance
for many experimental and theoretical applications to dissipative
reversible systems.}
\*

The hypothesis is a reinterpretation of a principle stated by Ruelle,
\ref{R3}{}{7}. Applications to microscopically irreversible systems
(like systems with microscopic friction or systems like the Navier
Stokes equations) have also been proposed, see \ref{G}{3}{8}.

In this paper we shall not deal with the applications (see \ref{ECM}{}{9} and
\ref{BGG}{}{10} for numerical simulations applications) but, rather, with
conceptual problems of interpretation of the fluctuation theorem and of
its extensions that we study below.
\*

\ifnum\mgnf=1 \pagina\fi
\0{\it\S2 Fluctuation patterns}
\numsec=2\numfor=1
\*

The fluctuation theorem can be interpreted, see above and \ref{G}{2}{5},
as extending to non equilibrium and to large fluctuations of time
averages, gaussian fluctuations theory and Onsager reciprocity.  Hence
it is natural to inquire whether there are more direct and physical
interpretations of the theorem (hence of the meaning of the chaotic
hypothesis) when the external forcing is really different from the value
$0$ (that is always assumed in Onsager's theory).  A result in this
direction is the {\it conditional reversibility theorem}, discussed
below.

Consider an observable $F$ which, for simplicity, has a well defined
time reversal parity: $F(Ix)=\e_F F(x)$, with $\e_F=\pm1$.  Let $F_+=0$
be its time average (\ie its SRB average) and let $t\to \f(t)$ be a
smooth function vanishing for $|t|$ large enough.  We look at the
probability, relative to the SRB distribution (\ie in the ``natural
stationary state'') that $F(S_t x)$ is $\f(t)$ for $t\in
[-\fra\t2,\fra\t2]$: we say that $F$ "follows the fluctuation pattern"
$\f$ in the time interval $t\in [-\fra\t2,\fra\t2]$.

No assumption on the fluctuation size, nor on the size of the
forces keeping the system out of equilibrium, will be made.  We assume,
however, that the evolution is time reversible {\it also} out of
equilibrium and that the phase space contraction rate $\s_+$ is not
zero (the results hold no matter how small $\s_+$ is and they make
sense even if $\s_+=0$, but they become trivial).

We denote $\z(p,\f)$ the {\it large deviation function} for observing
in the time interval $[-\fra\t2,\fra\t2]$ an average phase space
contraction $\s_\t\defi\fra1\t\ig_{-\t/2}^{\t/2}\s(S_tx)dt= p\s_+$ and
at the same time a fluctuation pattern $F(S_tx)=\f(t)$.  This means
that the probability that the {\it dimensionless average entropy
creation rate} $p$ is in a open set $\D$ and $F$ is in a
neighborhood\annota{1}{By "neighborhood" $U_{\V\psi,\e}$ we mean that
$\ig_{-\t/2}^{\t/2} \psi(t)F(S_tx)dt$ is approximated within
given $\e>0$ by $\ig_{-\t/2}^{\t/2} \psi(t)\f(t)dt$ for $\psi$ in the
finite collection $\V\psi=(\psi_1,\ldots,\psi_m)$ of test functions.}
$U_{\V \psi}$ of $\f$ is given by:

$$\sup_{p\in\D,\f\in U_{\V \psi}} e^{-\t\z_\t(p,\f)}\Eq(2.1)$$
to leading order as $\t\to\io$ (\ie the logarithm of the mentioned
probability divided by $\tau$ converges as $\t\to\io$ to
$\sup_{p\in\D,\f\in U_{\V \psi}} -\z(p,\f)$).

Given a reversible, dissipative, transitive Anosov flow the fluctuation
pattern $t\to\f(t)$ and the time reversed pattern $t\to\e_F\f(-t)$ are
then related by the following: \*

\0{\it Conditional reversibility theorem:} {\sl If $F$ is an observable
with defined time reversal parity $\e_F=\pm1$ and if $\t$ is large the
fluctuation pattern $\f(t)$ and its time reversal $I\f(t)\=\e_F\f(-t)$
will be followed with equal likelyhood if the first is conditioned to
an entropy creation rate $p$ and the second to the opposite $-p$.  This
means:

$$\fra{\z(-p,I\f)-\z(p,\f) }{p\s_+}=1 \qquad {\rm for\ } |p|\le
p^*\Eq(2.2)$$
with $\z$ introduced above and a suitable $p^*\ge1$.}
\*

In other words while it is very difficult, in the considered systems, that we
see an "anomalous" average entropy creation rate during a time $\t$
(\eg $p=-1$), it is also true that
``that is the hardest thing to see".  Once we see it
{\it all the observables will behave strangely} and the relative
probabilities of time reversed patterns will become as likely as those
of the corresponding direct patterns under "normal" average entropy
creation regime.

"A waterfall will go up, as likely as we see it going down, in a world
in which for some reason the entropy creation rate has changed sign
during a long enough time." We can also say that the motion on an
attractor is reversible, even in presence of dissipation, once the
dissipation is fixed.

The proof of the theorem is very simple and in fact it is a repetition
of the fluctuation theorem proof.  To be complete we sketch here
the proof of the corresponding result for discrete time systems, \ie
for systems whose evolution is a map $S$ of a smooth compact manifold
$\CC$, "phase space", and $S$ is a reversible map (\ie $IS=S^{-1}I$ for
a volume preserving diffeomorphism $I$ of phase space such that
$I^2=1$) and a dissipative, transitive Anosov map (see below).

The latter systems are simpler to study because Anosov maps do not have
a trivial Lyapunov exponent (the vanishing one associated with the
phase space flow direction); but the techniques to extend the analysis
to continuous time systems will be the same as those developed in
\ref{Ge}{}{4} for the similar problem of proving the fluctuation
theorem for Anosov flows.  We dedicate the following section to the
formulation of the conditional reversibility theorem for systems with
discrete time evolution.  \*

\0{\it\S3. The case of discrete time evolution}
\numsec=3\numfor=1
\*

A description of systems which evolve chaotically in discrete time is
possible closely following the one dedicated in the previous sections
to continuous time systems.

Consider a dynamical system described by a smooth map $S$ acting on a
smooth compact manifold $\CC$ and depending on a few parameters $\V G$
so that for $\V G=\V 0$ the system is volume preserving.  Suppose also
that the system is time reversible, \ie there is a volume preserving
diffeomorphism $I$ of $\CC$ which anticommutes with $S$: $IS=S^{-1}I$;
furthermore suppose that it is chaotic, \ie $(\CC,S)$ is a transitive
Anosov map.

We recall that an Anosov map on a smooth compact manifold $\CC$ is a
map such that at every point $y\in\CC$ one can define a {\it stable
tangent plane} $T^s_y$, and an {\it unstable tangent plane} $T^u_y$
with:\\
1) the tangent planes $T^u_y$ and $T^s_y$ are independent and vary
continuously with $y$ always spanning the full tangent plane at $y$.
\\
2) they are covariant $\dpr S T^\a_y=T^\a_{Sy}$, $\a=u,s$\\
3) there exist constants $C,\l>0$ such that if $\x\in T^s_y$ then
$|\dpr S^n \x|\le C e^{-\l n}|\x|$ for all $n\ge0$, and if $\x\in
T^u_y$ then $|\dpr S^n \x|\le C e^{-\l n}|\x|$ for all $n\ge0$.

A Anosov map is {\it mixing} or {\it transitive} if given any two open
sets $A,B$ there is a $n_{A,B}$ such that $S^n A\cap B\ne \emptyset$
for $|n|> n_{A,B}$.

If $(\CC,S)$ is a transitive Anosov system the time averages of
smooth observables on trajectories starting at almost all points, with
respect to the volume, do exist and can be computed as integrals over
phase space with respect to a probability distribution $\m$, which is
unique and is called the SRB distribution, \ref{S}{}{11}.

For reversible transitive Anosov maps a fluctuation theorem analogous to
the one for flows can be formulated as follows. Let $J(x)=\dpr S(x)$ be
the jacobian matrix of the transformation $S$.

The quantity $\lis\s(x)=-\log \L(x)$ will be called the {\it entropy
production} per timing event so that $e^{-\lis\s(x)}$ is the {\it phase
space volume contraction} per event: {\it it will be called entropy
creation rate per event}.  This is the analogue of the divergence of
the equations of motion in the continuous time systems considered in
\S1,2.

The non negativity of the time average, \ie of the SRB average, of
$\lis\s(x)$ is in fact a theorem that could be called the {\it
H--theorem} of reversible non equilibrium statistical mechanics,
\ref{R1}{}{12}.  It can be also shown, {\it c.f.r} \ref{R1}{}{12}, that the
average of $\lis\s(x)$ with respect to the SRB distribution can vanish
only if the latter has a density with respect to the volume.

Therefore we shall call {\it dissipative} systems for which the time
average of $\lis\s(x)$ is positive, \ref{GC}{}{3}.

We call $\lis\s_\t(x)$ the partial average of $\lis\s(x)$ over the part
of trajectory centered at $x$ (in time): $S^{-\t/2}x,\ldots,
S^{\t/2-1}x$.  Then we can define the {\it dimensionless entropy
production} $p=p(x)$ via:

$$\lis\s_\t(x)= \fra1\t\sum_{j=-\t/2}^{\t/2-1}\lis\s(S^j
x)\defi\media{\lis\s}_+ p\Eq(3.1)$$
where $\media{\lis\s}_+$ is the infinite time average $\ig_\CC
\lis\s(y)\m(dy)$, if $\m$ is the forward statistics of the volume
measure (\ie is the SRB distribution), and $\t$ is any integer.

Let $F$ be an observable with time reversal parity $\e_F=\pm1$ and let
$n\to \f(n)$ be a function vanishing for $|n|$ large enough, say $|n|>
n_0$.  If $\D, U_j$, $|j|< N$, are open intervals and $\t$ is given,
the probability that $p(x)\in \D, F(S^jx)\in U_j$ is given by:

$$\sup_{p\in\D, \f(j)\in U_j} e^{-\t\z(p,\f)}\Eq(3.2)$$
to leading order as $\t\to\io$ for some $\z(p,\f)$. And the following
theorem holds:
\*

\0{\it Conditional reversibility theorem} (discrete time case): {\sl
Given $F$ its fluctuation patterns $\f$, considered in a time interval
$[t-\fra12\t,t+\fra12\t]$ during which the entropy creation rate has
average $p$, have the same relative probability as their time reversed
patterns $I\f(n)=\e_F \f(-n)$ in a time interval
$[t'-\fra12\t,t'+\fra12\t]$ during which the entropy creation rate has
opposite average $-p$.}
\*

Analitycally this is again expressed by property \equ(2.2) of the
large deviation function $\z(p,\f)$.  The proof of this result is a
repetition of the proof of the fluctuation theorem of \ref{GC}{}{3}and
we sketch it in the following section for the case of reversible
dissipative transitive Anosov maps.

\*
\0{\it\S4. "Conditional reversibility" and "fluctuation"
theorems.}
\numsec=4\numfor=1
\*

We need a few well known, but sometimes quite non trivial, geometrical
results about Anosov maps.

The stable planes form an "integrable" family in the sense that there
are locally smooth manifolds having everywhere a stable plane as
tangent plane, and likewise the unstable manifolds are integrable.
This defines the notion of stable and unstable manifold through a
point $x\in\CC$: they will be denoted $W^s_x$ and $W^u_x$, see
\ref{R}{2}{13}.  Globally such manifolds wrap around in the phase space
$\CC$ and, in {\it transitive} systems, the stable and unstable
manifolds of each point are dense in $\CC$, see \ref{S}{}{11},
\ref{R}{2}{13}.

The stable and unstable planes covariance implies that if $J(x)=\dpr
S(x)$ is the jacobian matrix of $S$ we can regard its action mapping
the tangent plane $T_x$ onto $T_{Sx}$ as "split" linearly into an
action on the stable plane and one on the unstable plane: \ie $J(x)$
restricted to the stable plane becomes a linear map $J^s(x)$ mapping
$T^s_x$ to $T^s_{Sx}$. Likewise one can define the map $J^u(x)$,
\ref{R}{2}{13}.

We call $\L_u(x),\L_s(x)$ the determinants of the jacobians $J^u(x),
J^s(x)$: their product differs from the determinant $\L(x)$ of $\dpr
S(x)$ by the ratio of the sine of the angle $a(x)$ between the planes
$T^s_x,T^u_x$ and the sine of the angle $a(Sx)$ between
$T^s_{Sx},T^u_{Sx}$: hence $\L(x)=\fra{\sin a(Sx)}{\sin
a(x)}\L_s(x)\L_u(x)$.  We set also:

$$\eqalignno{
\L_{u,\t}(x)=&\prod_{j=-\t/2}^{\t/2-1}\L_u(S^jx),\quad
\L_{s,\t}(x)=\prod_{j=-\t/2}^{\t/2-1}\L_s(S^jx),\quad\cr
\L_{\t}(x)=&\prod_{j=-\t/2}^{\t/2-1}\L(S^jx)&\eq(4.1)\cr}$$

Time reversal symmetry implies that $W^s_{Ix}=IW^u_x, W^u_{Ix}=IW^s_x$
and:

$$\eqalign{
&\L_\t(x)=\L_\t(Ix)^{-1},\qquad \L_{s,\t}(Ix)=\L_{u,\t}(x)^{-1},
\qquad \L_{u,\t}(Ix)=\L_{s,\t}(x)^{-1}\cr
&\sin a(x)=\sin a(Ix)\cr}\Eq(4.2)$$

Given the above geometric--kinematical notions the SRB distribution
$\m$ can be represented by assigning suitable weights to small phase
space cells.  This is very similar to the representation of the
Maxwell--Boltzmann distributions of equilibrium states in terms of
suitable weights given to phase space cells of equal Liouville volume.

The phase space cells can be made consistently as small as we please
and, by taking them small enough, one can achieve an arbitrary
precision in the description of the SRB distribution  $\m$, in the same
way as we can approximate the Liouville volume by taking the phase space
cells small.

The key to the construction is a Markov partition: this is a partition
$\EE=(E_1,\ldots,E_\NN)$ of the phase space $\CC$ into $\NN$ cells which
are {\it covariant} with respect to the time evolution, see \ref{R}{2}{13},
in a sense that we do not attempt to specify here, and with respect to
time reversal in the sense that $I E_j=E_{j'}$ for some $j'$, see
\ref{G}{4}{14}.

Given a Markov partition $\EE$ we can "refine" it "consistently" as
much as we wish by considering the partition $\EE_T= \vee_{-T}^T
S^{-j}\EE$ whose cells are obtained by ``intersecting the cells of
$\EE$ and of its $S$--iterates; the cells of $\EE_T$ become
exponentially small with $T\to\io$ as a consequence of the
hypebolicity. In each $E_j\in\EE_T$ one can select a point $x_j$
("center": quite arbitrary and not uniquely defined, see
\ref{GC}{}{3}, \ref{G}{4}{14}) so that $Ix_j$ is the point
selected in $IE_j$. Then we evaluate the expansion rate
$\L_{u,2T}(x_j)$ of $S^{2T}$ as a map of the unstable manifold of
$S^{-T}x_j$ to that of $S^Tx_j$.

Using the elements $E_j\in\EE_T$ as cells we can define approximations
"as good as we wish" to the SRB distribution $\m$ because for all smooth
observables $F$ defined on $\CC$ it is:

$$\ig \m(dy) F(y)=\lim_{T\to\io}\ig m_T(dy) F(y)\defi
\lim_{T\to\io}\fra{\sum_{E_j\in \EE_T}  F(x_j)
\L_{u,2T}^{-1}(x_j)}{\sum_{E_j\in \EE_T}
\L_{u,2T}^{-1}(x_j)}\Eq(4.3)$$
where $m_T(dy)$ is implicitly defined here by the ratio in the {\it
r.h.s.} of \equ(4.3).  This deep theorem of Sinai, \ref{S}{}{11}, (and
\ref{BR}{}{1} in the continuous time case), see also \ref{GC}{}{3},
\ref{G}{4}{14}, is the basis of the technical part of our reversibility
theorem.

If $\D_a$ denotes an interval $[a,a+da]$ we first evaluate the
probability, with respect to $m_{\t/2}$ of \equ(4.3), of the event that
$a(x_j)=\lis\s_\t(x)/\media{\lis\s}_+\in \D_p$ and, also, that
$F(S^nx)\in\D_{\f_n}$ for $|n|< \t$, divided by the probability (with
respect to the same distribution) of the time reversed event that
$a(x)=\lis\s_\t(x)/\media{\lis\s}_+\in \D_{-p}$ and, also,
$F(S^nx)\in\D_{\e_F\f_{-n}}$ for $|n|< \t$; \ie we compare the
probability of a fluctuation pattern $\f$ in presence of average
dissipation $p$ and that of the time reversed pattern in presence of
average dissipation $-p$.  This is essentially:
$$\fra{\sum_{j,\,a(x_j)=p, F(S^nx_j)=\f(n)}
\L_{u,\t}^{-1}(x_j)}
{\sum_{j,\, a(x_j)=-p,
F(S^nx_j)=\e_F\f(-n)} \L_{u,\t}^{-1}(x_j)}\Eq(4.4)$$
Since $m_{\t/2}$ in \equ(4.3) is only an approximation to $\m_+$
{\it an error is involved in using} \equ(4.4) as a formula for the same
ratio computed by using the true $\m_+$ instead of $m_{\t/2}$.

It can be shown that this "first" approximation (among the two that
will be made) we can be estimated to affect the result
only by a factor bounded above and below uniformly in $\t,p$,
\ref{GC}{}{3}. This is {\it not} completely straightforward: in a sense
this is perhaps the main technical problem of the analysis: further
mathematical details can be found in \ref{GC}{}{3} and in \ref{Ge}{}{4}.

\*
{\it Remark:} There are other representations of the SRB distributions
that seem more appealing than the above one based on the (not too
familiar) Markov partitions notion.  The simplest is perhaps the {\it
periodic orbits representation} in which the role of the cells is taken
by the periodic orbits.  However I do not know a way of making the
argument that leads to \equ(4.4) {\it controlling} the approximations
which does not rely on Markov partitions.  And in fact I do not know of
any expression of the SRB distribution that is not proved by using the
very existence of Markov partitions.
\*
We now try to establish a one to one correspondence between the addends
in the numerator of \equ(4.4) and the ones in the denominator,
aiming at showing that corresponding addends have a {\it constant
ratio} which will, therefore, be the value of the ratio in \equ(4.4).

This is possible because of the reversibility property: it will be used
in the form of its consequences given by the relations \equ(4.2).

The ratio \equ(4.4) can therefore be written simply as:
$$\fra{\sum_{j,\,a(x_j)=p, F(S^nx_j)=\f(n)}\L_{u,\t}^{-1}(x_j)}
{\sum_{j,\,a(x_j)=-p, F(S^nx_j)=\e_F\f(-n)}\L_{u,\t}^{-1}(x_j)}
\=\fra{\sum_{j,\,a(x_j)=p, F(S^nx_j)=\f(n)} \L_{u,\t}^{-1}(x_j) }
{\sum_{j,\,a(x_j)=p, F(S^nx_j)=\f(n)} \L_{s,\t}(x_j)}\Eq(4.5)$$
where $x_j\in E_j$ is the center in $E_j$.  In deducing the second
relation we take into account time reversal symmetry $I$, that the
centers $x_{j}, x_{j'}$ of $E_j$ and $E_{j'}=IE_j$ are such that
$x_{j'}=Ix_j$, and \equ(4.2) in order to transform the sum in the
denominator pf the {\it l.h.s.} of \equ(4.5) into a sum over the same
set of labels that appear in the numerator sum.

It follows then that the ratios between corresponding terms in the ratio
\equ(4.5) is equal to $\L_{u,\t}^{-1}(x)\L_{s,\t}^{-1}(x)$.
This differs from the reciprocal of the total change of phase space
volume over the $\t$ time steps (during which
the system evolves from the point $S^{-\t/2}x$ to
$S^{\t/2 }x$).

The difference is only due to not taking into account the ratio of she
sines of the angles $a(S^{-\t/2}x)$ and $a(S^{\t/2}x)$ formed by the
stable and unstable manifolds at the points $S^{-\t/2}x$ and
$S^{\t/2}x$.  Therefore $\L_{u,\t}^{-1}(x)\L_{s,\t}^{-1}(x)$ will
differ from the actual phase space contraction under the action of
$S^\t$, regarded as a map between $S^{-\t/2}x$ and $S^{\t/2}x$, by a
factor that can be bounded between $B^{-1}$ and $B$ with
$B=\max_{x,x'}\fra{|\sin a(x)|}{|\sin a(x`)|}$, which is finite and
positive, by the linear independence of the stable and unstable
manifolds.

But for all the points $x_j$ in \equ(4.5), the reciprocal of the total
phase space volume change over a time $\t$ is
$e^{p\media{\lis\s}_+\t}$, which (by the constraint,
$\lis\s_\t/\media{\lis\s}_+=p$, imposed on the summation labels) equals
$e^{\t\media{\lis\s}_+\,p}$ up to a "second" approximation that cannot
exceed a factor $B^{\pm1}$ which is {\it $\t$-independent}.  Hence the
ratio \equ(4.4) will be the exponential: $e^{\t\media{\lis\s}_+\,p}$.

It is important to note that there have been two approximations, as
pointed out in the discussion above.  They can be estimated, see
\ref{GC}{}{3}, and imply that the argument of the exponential {\it is
correct up to $p,\f,\t$--independent corrections} making the
consideration of the limit $\t\to\io$ {\it necessary} in the formulation of
the theorem.
\*

\0{\it\S4. Concluding remarks.}
\numsec=4\numfor=1\*

1) One should note that in applications the above theorems will be used
through the chaotic hypothesis and therefore other errors may arise
because of its approximate validity (the hypothesis in fact
essentially states that "things go as if" the system was Anosov): they
may depend on the number $N$ of degrees of freedom and we do not
control them except for the fact that, if present, their relative value
should tend to $0$ as $N\to\io$: there may be (and very likely there
are) cases in which the chaotic hypotesis is not reasonable for small
$N$ (\eg systems like the Fermi-Pasta-Ulam chains) but it might be
correct for large $N$.  We also mention that, on the other hand, for
some systems with small $N$ the chaotic hypothesis may be already
regarded as valid (\eg for the models in \ref{CELS}{}{15},
\ref{ECM}{}{9}, \ref{BGG}{}{10}).

2) A frequent remark that is made about the chaotic hypothesis is that
it does not seem to keep the right viewpoint on nonequilibrium
thermodynamics. In fact it is analogous to the ergodic hypothesis which
({\it as well known}) cannot be taken as the foundation of equilibrium
statistical mechanics, even though it leads to the correct Maxwell
Boltzmann statistics, because the latter "holds for other reasons".
Namely it holds because in most of phase space (measuring sizes by the
Liouville measure) the few interesting macroscopic observable have the
same value, \ref{T}{}{16}, see also \ref{L}{}{17}.

An examination of the basic paper of Boltzmann \ref{B}{1}{18}, in
which the theory of equilibrium ensembles is developed, may offer some
arguments for further meditation.  The paper starts by illustrating
an important, and today almost forgotten, remark of Helmoltz showing
that very simple systems ("monocyclic systems") can be used to
construct mechanical models of thermodynamics: the example chosen by
Boltzmann is {\it really extreme by all standards}.  He shows that the
{\it Moon} motions on keplerian orbits can be used to build a model of
thermodynamics.

In the sense that one can call "volume" the radius of the orbit,
"temperature" the kinetic energy, "energy" the energy and "pressure"
another suitable kinematical quantity, associated with the orbital
parameters, and then one infers that by varying the orbit parameters (energy
and eccentricity) the relation $(dU+pdV)/T=\ {\it exact}$ holds.

Clearly this {\it could} be regarded as a curiosity.

However Boltzmann (following Helmoltz) took it seriously and proceeded
to infer that under the ergodic hypothesis {\it any system} small or
large provided us with a model of thermodynamics (as it was monocyclic
in the sense of Helmoltz): for instance he showed that the canonical
ensemble verifies exactly the second principle of thermodynamics (in
the form $(dU+pdV)/T=\ {\it exact}$) {\it without any need to take
thermodynamic limits}, \ref{B}{1}{18}, \ref{G}{5}{19}.  The same could
be said of the microcanonical ensemble (here, however, he had to change
"slightly" the definition of heat to make things work without finite
size corrections).

He realized that the ergodic hypothesis could not possibly
account for the correctness of the canonical (or microcanonical)
ensembles: this is clear at least from his (later) paper in defense
against Zermelo's criticism, \ref{B}{2}{20}, nor for the observed time
scales of approach to equilibrium.  Nevertheless he called
the theorem he had proved the {\it heat theorem} and never seemed to
doubt that it provided evidence for the correctness of the use of the
equilibrium ensembles for equilibrium statistical mechanics.

Hence we see that there are two aspects to consider: first certain
relations among mechanical quantities {\it hold no matter how large}
is the size of the system and, secondly, they can be seen and tested
even in large systems because in large systems such mechanical
quantities acquire a macroscopic thermodynamic meaning and their
relations are "typical" \ie they hold in most of phase space.

The consequences of the chaotic hypothesis stem from the properties of
small Anosov systems (the best understood) which play here the role of
Helmoltz's monocyclic systems of which Boltzmann's Moon is the
paradigm.  They are remarkable consequences because they provide us
with {\it parameter free relations} (namely the fluctuation theorem and
its consequences): but clearly we cannot hope to found solely upon them
a theory of nonequilibrium statistical mechanics because of the same
reasons the validity of the second law for monocyclic systems had in
priciple no reason to imply the theory of ensembles.

Thus what is missing are arguments similar to those used by Boltzmann
to justify the use of the ensembles independently of the ergodic
hypothesis: an hypothesis which in the end may appear (and still does
appear to many) as having only suggested them "by accident".  The
missing arguments should justify the fluctutation theorem on the basis
of the extreme likelihood of its predictions in systems that are very
large and that may be not Anosov systems in the mathematical sense.  I
see no reason, now, why this should prove impossible {\it a priori} or in
the future.

In the meantime it seems interesting to take the same philosophical
viewpoint adopted by Boltzmann: not to consider a chance that {\it all}
chaotic systems share some selected properties and try to see if such
properties help us achieving a better understanding of
nonequilibrium. After all it seems that Boltmann himself took a rather
long time to realize the interplay of the above two basic mechanisms
behind the equilibrium ensembles and to propose a solution harmonizing
them. "All it remains to do" is to explore if the hypothesis has
implications more interesting or deeper than the fluctuation theorem,
see \ref{G}{3}{8}, \ref{GR}{}{6}.  \*

\0{\it Acknowledgements:} I are indebted to F.  Bonetto, E.G.D.  Cohen,
J.  Lebowitz, D.  Ruelle for stimulating discussions and comments.
This work is part of the research program of the European Network on:
"Stability and Universality in Classical Mechanics", \#
ERBCHRXCT940460.

\*
{\bf References}
\*

\def\aB{Boltzmann, L.: {\it \"Uber die eigenshaften monzyklischer
und anderer damit verwandter Systeme}, in "Wissenshafltliche
Abhandlungen", ed. F.P. Hasenh\"orl, vol. III,
Chelsea, New York, 1968, (english, 1884 original).}
\rif{B}{1}{\aB}{0}

\def\bB{ Boltzmann, L.: {\it Entgegnung auf die w\"armetheoretischen
Betrachtungen des Hrn.  E.  Zermelo}, engl.  transl.: S.  Brush,
"Kinetic Theory", {vol. 2}, 218, Pergamon Press, 1966. And: {\it Zu Hrn.
Zermelo's Abhandlung "Ueber die mechanische Er\-kl\"a\-rung
irreversibler Vorg\"ange}, engl.  trans. in S.  Brush, "Kinetic Theory",
{\bf 2}, 238, (english,  1896 and 1897 originals).}
\rif{B}{2}{\bB}{0}

\def\aBR{Bowen, R., Ruelle, D.: {\it Axiom A flows}, Inventiones
Mathematicae, 197?}
\rif{BR}{}{\aBR}{0}

\def\aBGG{ Bonetto, F., Gallavotti, G., Garrido, P.: {\it Chaotic
principle: an experimental test}, mp$\_$arc @math. utexas. edu,
\# 96-154, in print on Physica D.}
\rif{BGG}{}{\aBGG}{}

\def\aCELS{ Chernov, N. I., Eyink, G. L., Lebowitz, J.L., Sinai, Y.:
{\it Steady state electric conductivity in the periodic Lorentz gas},
Communications in Mathematical Physics, {\bf 154}, 569--601, 1993.}
\rif{CELS}{}{\aCELS}{0}

\def\aECM{ Evans, D.J.,Cohen, E.G.D., Morriss, G.P.: {\it Probability
of second law violations in shearing steady flows}, Physical Review
Letters, {\bf 71}, 2401--2404, 1993.}
\rif{ECM}{}{\aECM}{0}

\def\aG{Gallavotti, G.: {\it Chaotic hypothesis: Onsager reciprocity and
fluctuation--dissi\-pa\-tion theorem}, Journal of Statistical Physics, {\bf
84}, 899--926, 1996. See also: {\it Chaotic principle: some
applications to developed turbulence}, in {\it
mp$\_$arc@\-math.\-utexas.\-edu}, \#95-232, and {\it chao-dyn@ xyz.lanl.gov},
\#9505013, in print in J. of Statistical Physics.}
\rif{G}{1}{\aG}{0}

\def\bG{Gallavotti, G.: {\it Extension of Onsager's reciprocity to
large fields and the chao\-tic hypothesis}, Physical Review
Letters, {\bf 78}, 4334--4337, 1996.}
\rif{G}{2}{\bG}{0}

\def\cG{Gallavotti, G.: {\it Equivalence of dynamical ensembles and
Navier-Stokes equations}, Phy\-sics Letters, {\bf223}, 91--95, 1996.  And
{\it Dynamical ensembles equivalence in statistical mechanics}, in
mp$\_$arc@ math.  utexas.  edu \#96-182, chao-dyn@xyz.lanl.gov
\#9605006, in print on Physica D.}
\rif{G}{3}{\cG}{0}

\def\dG{Gallavotti, G.: {\it New methods in nonequilibrium
gases and fluids}, in mp$\_$arc@ math. utexas.edu \#96-533,
and chao-dyn \#9610018.}
\rif{G}{4}{\dG}{0}

\def\eG{Gallavotti, G.: {\it Trattatello di Meccanica Statistica},
Quaderni CNR-GNFM, {\bf 50}, p.  1--350, 1995, Firenze. Free copies of
the book (written 70\% italian and 30\%english) can be obtained by
requesting them to CNR--GNFM, via S. Marta 13/a, 50139 Firenze, Italy,
or by fax at CNR-GNFM -39-55-475915.}  \rif{G}{5}{\eG}{0}

\def\fG{Gallavotti, G.: {\it Reversible Anosov maps and large deviations},
Mathematical Physics Electronic Journal, MPEJ, (http:// mpej.unige.ch),
{\bf 1}, 1--12, 1995.}
\rif{G}{6}{\fG}{0}

\def\aGe{Gentile, G.: {\it Large deviation rule for Anosov flows}, IHES
preprint, in print in Forum Mathematicum.}
\rif{Ge}{}{\aGe}{0}

\def\aGC{Gallavotti, G., Cohen, E.G.D.: {\it Dynamical ensembles in
nonequilibrium statistical mechanics}, Physical Review Letters, {\bf74},
2694--2697, 1995; {\it Dynamical ensembles in stationary states},
Journal of Statistical Physics, {\bf 80}, 931--970, 1995. See also:
\ref{G}{6}{2}.}
\rif{GC}{}{\aGC}{0}

\def\aGR{Gallavotti, G., Ruelle, D.: {\it SRB states and nonequilibrium
statistical mechanics close to equilibrium}, mp$\_$arc@math. utexas. edu,
\#96-645.}
\rif{GR}{}{\aGR}{0}

\def\aL{Lebowitz, J.L.: {\it Microscopic reversibility and macroscopic
behavior: physical explanations and mathematical derivations},
To appear in {\sl Enciclopedia del Novecento}, Enciclopedia Italiana,
Roma, 1997, english preprint: Rutgers University, 1995.}
\rif{L}{}{\aL}{0}

\def\aR{Ruelle, D.: {\it Positivity of entropy production in
nonequilibrium statistical mechanics}, Journal of Statistical Physics,
{\bf 85}, 1--25, 1996.  1996.}
\rif{R}{1}{\aR}{0}

\def\bR{Ruelle, D.: {\it Elements of differentiable dynamics and
bifurcation theory}, Academic Press, 1989.}
\rif{R}{2}{\bR}{0}

\def\cR{Ruelle, D.: {\it Chaotic motions and strange attractors},
Lezioni Lincee, notes by S.  Isola, Accademia Nazionale dei Lincei,
Cambridge University Press, 1989; see also: Ruelle, D.: {\it Measures
describing a turbulent flow}, Annals of the New York Academy of
Sciences, {\bf 357}, 1--9, 1980.}
\rif{R}{3}{\cR}{0}

\def\aS{Sinai, Y.G.: {\sl Gibbs measures in ergodic theory},
Russian Mathematical Surveys, {\bf 27}, 21--69, 1972.  Also: {\it
Introduction to ergodic theory}, Prin\-ce\-ton U.  Press, Princeton,
1977.}
\rif{S}{}{\aS}{0}

\def\aT{Taylor, W.: {\it The kinetic theory of the dissipation of energy},
reprinted in: S. Brush, {\it Kinetic theory}, {\bf 2}, 176,
Pergamon Press, 1966.}
\rif{T}{}{\aT}{0}

\ifnum\bibl=1 \input fin \fi

\FINE

\end